*Article*

# A New Agent-Based Methodology for the Seismic Vulnerability Assessment of Urban Areas

**Annalisa Greco** [1,*]**, Alessandro Pluchino** [2]**, Luca Barbarossa** [1]**, Giovanni Barreca** [3]**, Ivo Caliò** [1]**, Francesco Martinico** [1] **and Andrea Rapisarda** [2,4]

[1]  Department of Civil Engineering and Architecture, University of Catania, via S. Sofia 64, 95123 Catania, Italy; luca.barbarossa@darc.unict.it (L.B.); icalio@dica.unict.it (I.C.); francesco.martinico@unict.it (F.M.)
[2]  Department of Physics and Astronomy "Ettore Majorana", University of Catania, and INFN Sezione di Catania, via S. Sofia 64, 95123 Catania, Italy; alessandro.pluchino@ct.infn.it (A.P.); andrea.rapisarda@ct.infn.it (A.R.)
[3]  Department of Biological, Geological and Environment Science, University of Catania, Corso Italia 57, 95129 Catania, Italy; g.barreca@unict.it
[4]  Complexity Science Hub, Josefstaedter Strasse 39, 1080 Vienna. Austria.
*  Correspondence: annalisa.greco@unict.it; Tel.: +39-095-7382251

**Abstract:** In order to estimate the seismic vulnerability of a densely populated urban area; it would in principle be necessary to evaluate the dynamic behaviour of individual and aggregate buildings. These detailed seismic analyses, however, are extremely cost-intensive and require great processing time and expertise judgment. The aim of the present study is to propose a new methodology able to combine information and tools coming from different scientific fields in order to reproduce the effects of a seismic input in urban areas with known geological features and to estimate the entity of the damages caused on existing buildings. In particular, we present new software called ABES (Agent-Based Earthquake Simulator), based on a Self-Organized Criticality framework, which allows to evaluate the effects of a sequence of seismic events on a certain large urban area during a given interval of time. The integration of Geographic Information System (GIS) data sets, concerning both geological and urban information about the territory of Avola (Italy), allows performing a parametric study of these effects on a real context as a case study. The proposed new approach could be very useful in estimating the seismic vulnerability and defining planning strategies for seismic risk reduction in large urban areas

## 1. Introduction

The evaluation of the seismic vulnerability of existing buildings has been extensively studied during the last 30 years at different scales, from the dimension of a single building to large urban areas. A reliable vulnerability evaluation for a single building requires expert analytical calculations and a deep knowledge of the geometry of the structure, of its mechanical properties and of the characteristic parameters of the foundation soil. It is evident that, due to the amount of data and resources involved in a rigorous assessment, it is economically unsustainable to extend to large urban contexts the detailed analyses developed on each single building. The change in scale involves therefore a reduction in the accuracy of the results. Nevertheless, in order to define planning strategies for the reduction of seismic risk at urban scale, it is very important to be able to perform vulnerability assessments, based on simplified approaches and rapid processing.

Several procedures for a synthetic assessment of the seismic vulnerability of aggregates of buildings representing portions of urban areas have been already presented in the scientific literature [1–12]. In particular, among the others: Ramos and Lourenço [1] studied the vulnerability of the masonry buildings in the historical city center of Lisbon through a finite element method; Senaldi et al. [3] focused on the seismic response of masonry building aggregates obtained by means of non linear dynamic analysis; Giovinazzi and Pampanin [10] proposed a simple, but reliable approach to assess, on different scale levels, the beneficial impact of seismic retrofitting for reinforced concrete buildings built before 1970; Greco et al. [11] applied an innovative macro-element approach for the assessment of the seismic vulnerability of typical unreinforced masonry buildings in the area of Catania (Italy); Lestuzzi et al. [12] approached the seismic assessment at urban scale of the cities of Sion and Martigny (Switzerland) adopting the Risk-UE methodology, in particular the empirical and mechanical methods (LM1 and LM2).

Seismic vulnerability assessments are referred to the expected seismic actions on the specific site, which can only be statistically presumed from previous recorded data and on the basis of the geological characteristic. Prescriptions of national codes refer to the seismic hazard of the construction site, that depends on the elastic response spectra related to the soil typology and to the peak ground acceleration, which can be exceeded with a predetermined probability in the reference period of the construction. The seismic data related to the various construction sites are scaled according to the maximum acceleration expected on the ground, which, on the basis of past events, is related to the possibility of occurrence of a single seismic event of a certain intensity. Anyway, large seismic events are often not isolated but are preceded and followed by a foreshock and an aftershock activity of variable intensity and duration (Omori Law) [13–16]. For example, the severe earthquake of magnitude 5.9 ML (Richter Scale) occurred in L'Aquila (Italy) on April 6 2009, at 3:32 a.m. (that caused more than 305 victims, 1500 wounded, 7000 people homeless and 4 bn euros of estimated damages to the buildings), was the mainshock of an anomalous activity which started in December 2008 and lasted until 2012 [17]. In order to give an idea of the great number of shocks involved, it is interesting to highlight that just in the year that followed the April 6 event, the Italian agency INGV (Istituto Nazionale di Geofisica e Vulcanologia, www.ingv.it) reported that about 18,000 earthquakes occurred only across the area of the city of L'Aquila with different epicenters (256 events were registered only during the 48 hours immediately after the mainshock, 56 of them with a magnitude greater than 3 ML).

Apart from such catastrophic seismic scenarios, moderate ground activities are recorded every day all over the earth. When the intensity of the seismic input exceeds a minimum value, related to the structural characteristic of each building, the latter can suffer some damage, that in some cases could be difficult to identify but leads to a reduction of the seismic resistance of the structure. Therefore, not only severe ground motions constitute a danger for structures since damage can occur even for moderate seismic actions and a building can collapse after several small earthquakes due to an incremental cumulative damage. In a seismic impact evaluation at regional or urban scale it would be very useful to have the possibility to estimate both the collapse scenario under severe earthquakes and the cumulative one caused by moderate and repetitive ground shakings. Of course, the estimation of the seismic vulnerability of a given urban area strongly depends on the characteristics of the seismic input that can be simulated by means of modern computational techniques possibly supported by Geographic Information System data sets. Different approaches for the simulation of the seismic input have been recently presented by many researchers [18–22].

In particular, just to give a few examples: Lu et al. [18] proposed a coarse-grained parallel approach for the simulation of the seismic damage in urban areas based on refined models and GPU/CPU cooperative computing; Xiong et al. [19] introduced a nonlinear multiple degree-of-freedom flexural-shear model to better predict the responses of tall buildings in regional seismic simulations; Matassoni et al. [20] applied a Decision Support System (DSS) to the simulation of the seismic and impact scenarios for two major historical earthquakes recorded in Florence (Italy); Silva and Horspool [21] used a new methodology that combines the available information from the USGS ShakeMap system with the open-source software OpenQuake engine in order to estimate the number of structural collapses or economic losses after two different real earthquakes in Italy and New Zeland; Hancilar et al. [22] developed a new software, called ELER-Earthquake Loss Estimation Routine, for a rapid estimation of earthquake shaking and losses throughout the Euro-Mediterranean region.

In this paper we propose an innovative methodology that integrates different tools and sources of information in order to investigate the possible effects of a sequence of seismic events on the urban settlement of a given geographical area, assuming that this area is in the so-called "critical state". The concept of critical state has been developed in the context of the Self-Organized Criticality (SOC) theory, introduced in 1987 by Bak, Tang and Wiesenfeld [23]. SOC theory states that many large interactive systems observed in nature can self-organize into the critical state [24], a particular condition in which small perturbations may result in chain reactions which can affect any number of elements within the system.

Among many other fields, the SOC dynamics has been successfully applied also to seismology. In particular, in 1992 Olami-Feder-Christensen (OFC) realized a simple model which mimics a portion of terrestrial crust in the critical state [25]. Despite its intrinsic limitations (for example the depth of the epicenter and the features of the seismic waves are neglected), a dissipative version of the OFC model on a regular square lattice with a few long-range interactions has shown to be able to reproduce, with a good degree of approximation, the scale-invariant dynamics of real earthquakes [26]. As a matter of fact, when a given seismic area enters, after a given transient, into a critical state, the average earthquakes activity increases and events of any scale may occur. This is probably what happened between 2008 and 2012 in the territory of L'Aquila (Italy): The region entered into a critical state, and therefore the probability of occurrence of a large earthquake, like that one of April 6 2009, was no more negligible, even if—as a consequence of the SOC dynamics—it would have been impossible to predict the exact moment in which that event would have been realized.

The methodology here proposed adopts the SOC earthquake engine of the OFC model in the context of agent-based simulations. Agent-based (or multi-agent) simulations is a powerful computational tool that has been extensively applied in several fields to model complex phenomena [27–36]. This approach has been here implemented by means of a new software, called ABES (Agent-Based Earthquake Simulator) and realized within the programmable multi-agent environment NetLogo [37], with the purpose of assessing the seismic vulnerability of urban areas in the critical state. In order to take into account the structural characteristics of the existing buildings and the geological morphology of the territory considered as case study, the ABES simulation tool has been also integrated, within the same NetLogo environment, with real information taken from Geographic Information System (GIS) data sets. Unlike the other already cited studies, which typically focus on the effects produced on a given urban area by single severe seismic events, our analysis assumes that the crust below the considered area experiences a long sequence of earthquakes of any size with epicenters located in different parts of the considered territory. Therefore we are able to simulate the increase of buildings' damage due to repetitive ground motions.

This methodology is very general and could be applied to areas of any size (being the SOC approach self-similar and scale invariant) but, in order to show its effectiveness, we have chosen as a case study the territory around Avola (Siracusa), a small city in the southeast part of Sicily, for which both urban and geological GIS data are available. This zone, from the point of view of the seismic risk, is very similar to the area around L'Aquila. This will allow us to calibrate the ABES software in order to reproduce a damage scenario analogous to the one observed in L'Aquila region in 2009. Then, we will address new seismic scenarios into the critical state and explore the possible effects of different earthquakes sequences on the existing buildings, also highlighting the influence of some characteristic parameters of the buildings and of the soil on the occurrence of damage.

## 2. Materials and Methods

### 2.1. Seismic Vulnerability and Damage Evolution in Existing Buildings

The evaluation of a synthetic vulnerability value for each building in an urban area must take into account several parameters which, among others, consider the structural geometry, its age, the mechanical properties of the material, the quality of the construction and the geological characteristics of the site. A reliable estimate of the seismic vulnerability of a single existing building needs therefore a significant amount of data even for a synthetic appraisal.

In absence of sufficient information, a representative vulnerability index related to the structural typology of each building can anyway be assumed following some approximate approaches presented in the scientific literature. An interesting macroseismic model for the vulnerability assessment of existing buildings can be found in [2], where suitable ranges $V_{min} - V_{max}$ for the vulnerability of masonry and reinforced concrete typologies, are presented. These two construction typologies are almost the only ones present in Sicily and in particular in the territory of Avola and will therefore be considered in the applicative section. A summary of the considered ranges for the vulnerability index $V$ is reported in Table 1 (adapted from Ref.[2]) and it can be observed that it varies from a minimum value $V_{min} = -0.02$ for structures with high earthquake resistant design (E.R.D) to a maximum value $V_{max} = 1.02$ in total absence of E.R.D. As explained in Section 2.3, these ranges will be adopted in the present paper to assign the "initial vulnerability" $V_0$ to each building, following also the information contained in the urban GIS data set. This vulnerability will be successively updated taking into account the damage produced by the seismic ground motion intensity.

**Table 1.** Reference vulnerability values for building typologies (adapted from [2]).

| Typologies | Building type | $V_{min}$ | $V_{max}$ |
|---|---|---|---|
| Masonry | Rubble stone and earth bricks | 0.62 | 1.02 |
| | Simple stone | 0.46 | 1.02 |
| | Massive stone | 0.3 | 0.86 |
| | Masonry with old bricks | 0.46 | 1.02 |
| | Masonry with r.c. floors | 0.3 | 0.86 |
| | Reinforced /confined masonry | 0.14 | 0.7 |
| Reinforced Concrete | Frame in r.c. (without E.R.D) | 0.3 | 1.02 |
| | Frame in r.c. (moderate E.R.D.) | 0.14 | 0.86 |
| | Frame in r.c. (high E.R.D.) | -0.02 | 0.7 |
| | Shear walls (without E.R.D) | 0.3 | 0.86 |
| | Shear walls (moderate E.R.D.) | 0.14 | 0.7 |
| | Shear walls (high E.R.D.) | -0.02 | 0.54 |

For the intensity of the seismic input, the classifications used in the European Macroseismic Scale (EMS-98) [38], with its 12 levels of increasing damage, is here adopted: I. Not felt, II. Scarcely felt, III. Weak, IV. Largely observed, V. Strong, VI. Slightly damaging, VII. Damaging, VIII. Heavily damaging, IX. Destructive, X. Very destructive, XI. Devastating, XII. Completely devastating. This classification can be related to the most commonly adopted earthquake intensity scales, as shown in [38]. In particular, the macroseismic intensity is considered as a continuous parameter in the range 1-12 evaluated taking also into account possible amplification effects, with respect to rigid soil condition, depending on the mechanical characteristics of the site. This means that the intensity of an earthquake, with a given seismic magnitude and a given released energy, can be perceived by the buildings differently in different areas according to the corresponding geological typologies.

In the present paper the expected damage $\mu_D$ on each building has been related to the seismic input by means of a closed analytical function provided in the literature [2], which has been derived from EMS-98 macroseismic scale and also verified and calibrated on real damage data from different earthquakes. This function has the following sigmoidal expression:

$$\mu_D[I(M,c)] = 2.5\left[1 + \tanh\left(\frac{I(M,c) + 6.25V - 13.1}{Q}\right)\right] \qquad (1)$$

where $I(M,c)$ is the macroseismic intensity expressed as function of both the magnitude M of the earthquake and the parameter *c*, which represents the amplification coefficient of the soil below the building, while $V$ and $Q$ are, respectively, building's vulnerability and ductility indexes. Following Ref.[2], the value $Q = 2.3$ has been assumed for the ductility index of masonry buildings, judged to be representative for buildings not specifically designed to have ductile behaviour. Most of the reinforced concrete buildings have been designed without taking into account the earthquake loadings, therefore for all the r.c. building in the case study, a ductility index $Q = 2.6$ has been assumed.

In this paper, aiming at considering the damage cumulative process related to repetitive events, a simple original strategy for evaluating the reduction of structural performance associated to a sequence of earthquakes is proposed. At a certain state, the total damage $\mu_D^{TOT} = \sum \mu_D$ for each building is defined as the sum of the damage parameters $\mu_D$ for each previous seismic event, evaluated according to Equation (1). At the same time, the current vulnerability $V_{new}$ is assumed to follow the update rule:

$$V_{new} = V_0\left(1 + \frac{\mu_D^{TOT}}{5}\right) \qquad (2)$$

where $V_0$ is the initial vulnerability, assigned according to Table 1 (and also to Table 2, see later). This means that subsequent earthquakes can progressively injure undamaged buildings, which in turn increase their total damage $\mu_D^{TOT}$ (which starts from 0 at t=0) and change their status according to the value of $\mu_D^{TOT}$. In particular, following a classification reported in [2], when $0.5 \leq \mu_D^{TOT} < 2$ a given building changes its status from undamaged to "slightly (or moderately) damaged", when $2 \leq \mu_D^{TOT} < 4$ the same building results to be "heavily (or very heavily) damaged", and when $4 \leq \mu_D^{TOT} \leq 5$ it becomes "destroyed".

The evaluation of the total damage for each building can be very useful since it allows to globally visualize at the urban scale the areas with the same level of damage after each seismic input.

## 2.2. OFC: A Self-Organized Criticality Model of Earthquakes

The possibility of predicting earthquakes is a very old and debated problem which has stimulated many investigations in the last decades. As explained in the introduction, one of the most realistic models, able to mimic the seismic activity dynamics, was proposed by Olami-Feder-Christensen (OFC) within the framework of Self-Organized Criticality. In this context, in ref. [25] it has been shown that it is possible to reproduce the statistical features of different earthquakes catalogues by adopting a modified version of the OFC model.

The OFC model, which is at the basis of the ABES software presented in this paper, can be viewed as a two-dimensional square lattice of side *L* with *N* sites. A seismogenic force $F_i$ (seismic stress) acts on each site, which is connected to its four nearest neighbours. This force is a real number in the range $[0, F_{th}]$. To model a uniform tectonic loading dynamics as a function of time, all the forces are increased simultaneously and uniformly until one of them reaches the threshold value $F_{th}$ (typically $F_{th} = 1$) and becomes "active". At this point, the loading stops and an "earthquake" can start: The active node transfers a fraction $\alpha$ of its force to the four neighbours, which can in turn become active and pass the force to other neighbours, and so on and so forth. This simple dynamical rule can be written as

$$F_i \geq F_{th} \rightarrow \begin{cases} F_i \rightarrow 0 \\ F_{jj} \rightarrow F_{jj} + \alpha F_i \end{cases} \tag{3}$$

where "$jj$" denotes the set of nearest-neighbour nodes of $i$. The size S of a given earthquake, which represents the energy released by the seismic event, is given by the total number of sites activated during the dynamics. The parameter $\alpha$ controls the dissipation: The model is conservative if $\alpha = 0.25$, while it is dissipative for $\alpha < 0.25$. In Figure 1 a sketch of the dynamical rules of the OFC model is reported in order to clarify the earthquakes' formation process.

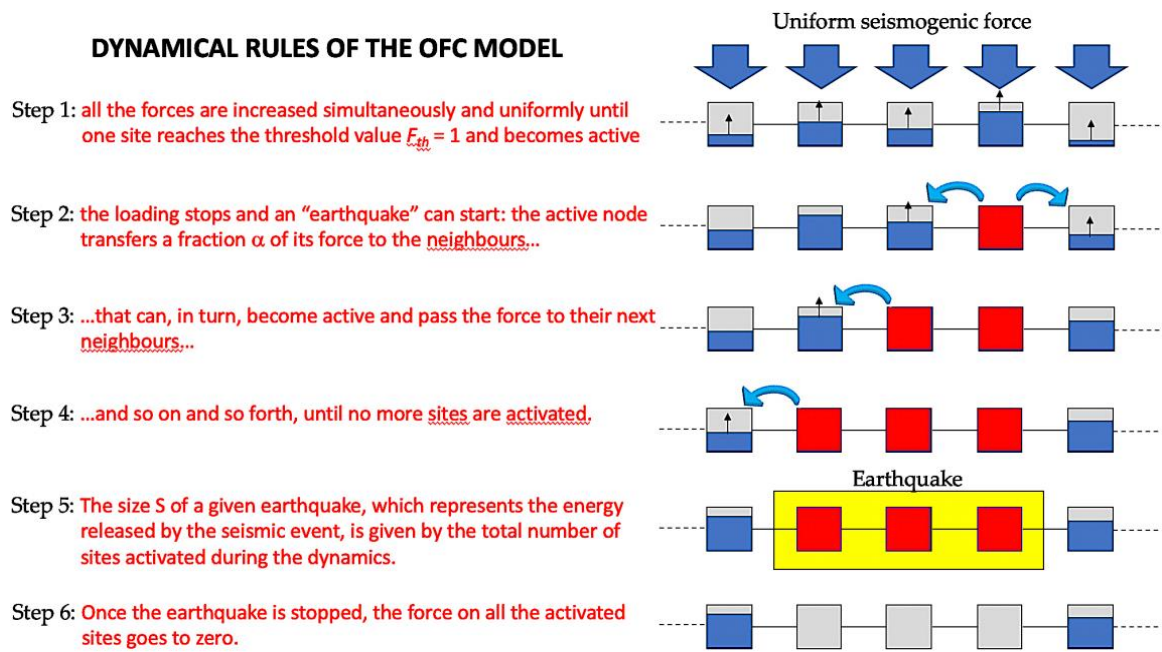

**Figure 1.** A sketch of the dynamical rules of the Olami-Feder-Christensen (OFC) model.

The modification of the OFC model proposed in [25] did introduce long-range correlations in the original lattice, therefore transforming it in a small world graph, a topological structure very common for many real networks characterized by local clustering and a short average distance among its nodes [39]. Actually, the presence of just a few long-range links seems able to better simulate the features of real seismic faults, by creating shortcuts that connect sites (nodes) which otherwise would be much further apart. As it has been shown, this kind of structure facilitates the system synchronization and produces both finite-size scaling and universal scaling exponents.

In this paper we adopt the small world version of the dissipative OFC model, with $\alpha = 0.21$. In particular, the model is implemented on a regular grid network 40 x 40 with a total N = 1600 nodes, where the links are rewired at random with a small probability p = 0.02 (typical of small world networks). Open boundary conditions are considered, i.e.; $F_i = 0$ on the boundary nodes.

The resulting network is shown in Figure 2(a), where nodes in red represent the S sites activated by a generic earthquake of size S. In the top panel of Figure 2(b) the size S of 2000 subsequent earthquakes during a typical run of the OFC dynamics, implemented by the ABES software, is plotted. After a transient of about 600 events, where the maximum size involves less than 5% of the entire lattice, the system enters into a critical state, where the average size of the earthquakes starts to increase and large events, involving a great number of nodes, have a non-zero probability of occurrence. The presence of criticality in the earthquakes' sequence is revealed by the probability distribution function (pdf) of the size S, which can be well fitted by a power-law with slope −1.72 (dashed straight line) in the log-log plot shown in the bottom panel of Figure 2(b). The power-law is also the signature of a scale-invariant behaviour, meaning that the size distribution of the earthquakes has a self-similar structure at all spatial scales.

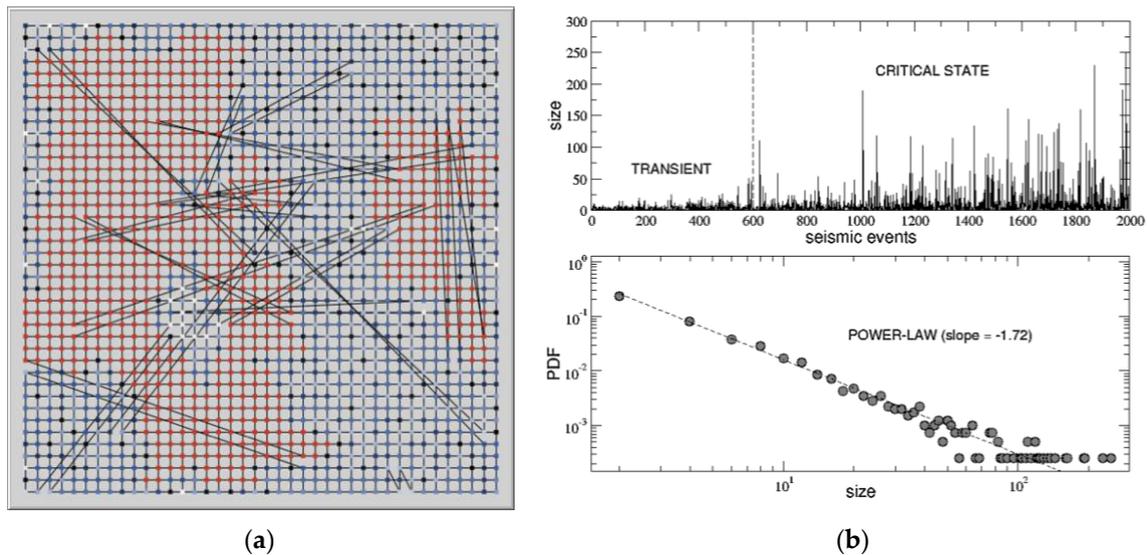

**(a)** **(b)**

**Figure 2.** (**a**) The small world lattice of the OFC model; (**b**) The sequence of earthquakes' sizes in the transient and in the critical state (upper panel) with the corresponding probability distribution (lower panel).

In order to adapt the OFC model output to the classification used in the European Macroseismic Scale (EMS-98), one needs to transform the size S (i.e.; the energy released) of a given earthquake into the corresponding intensity *I*, which—as already said—presents 12 different possible levels. The first step is to calculate the magnitude *M* of the earthquake, which is usually defined as the natural logarithm of the released energy: $M = ln\ S$ (since the energy released is an exponential function of the magnitude: $S = e^M$). Then, the magnitude can be transformed in the macroseismic intensity through the empirical relation $I(M) = 1.71\ M - 1.02$, obtained through a comparison between the magnitude scale and the EMS-98 one [40]. For example, the first noticeable peak (after the transient) in the sequence shown in the top panel of Figure 2(b) has a size $S \cong 100$ nodes, then a magnitude $M \cong 4.6$ and an intensity $I(M) \cong 6.85$. Finally, due to the geological characteristic of the soil, the final intensity perceived by the buildings in a given area (and adopted in Equation (1)) will be:

$$I(M,c) = I(M) + c \tag{4}$$

where *c* is the previously introduced amplification index characteristic of that area.

## 2.3. The Case Study of Avola and Description of the Urban GIS Data Set

In this section, the case study of Avola is introduced and the integration within the ABES software of two, urban and geological, GIS data sets is discussed. It is worth to stress that this has to be considered only as an example finalized to show the effectiveness of the proposed methodology in providing an estimation of seismic vulnerability of a given urban area.

The city of Avola (31576 inhabitants in 2016) is located along the south-east coast of Sicily, the so-called Val di Noto, thirty kilometres south of Siracusa. According to the Italian seismic hazard map (http://zonesismiche.mi.ingv.it), this area is very similar to the one around L'Aquila. The maximum acceleration expected (PGA peak ground acceleration) on the analysed territory is between 0.200 and 0.225 g, with probability of exceeding 10% in 50 years (according to "Progetto DPC-INGV-S1", see: http://esse1-gis.mi.ingv.it)

Avola was completely destroyed in 1693 by a major earthquake that hit South-eastern Sicily, causing thousands of victims. This catastrophic event caused a complete change in the structure of the entire 'Val di Noto' area, where a number of cities were rebuilt in new sites, closer to the coast.

After the earthquake, also the city of Avola was rebuilt in a different site according to a completely new layout in the coastal plain, one kilometre far from the coastline. The urban structure is characterized by a grid of perpendicular streets within a hexagonal perimeter (see Figure 3). A large main square, with nearby minor ones, marks the heart of the town, according to a design inspired by the ideal cities plans from the Renaissance.

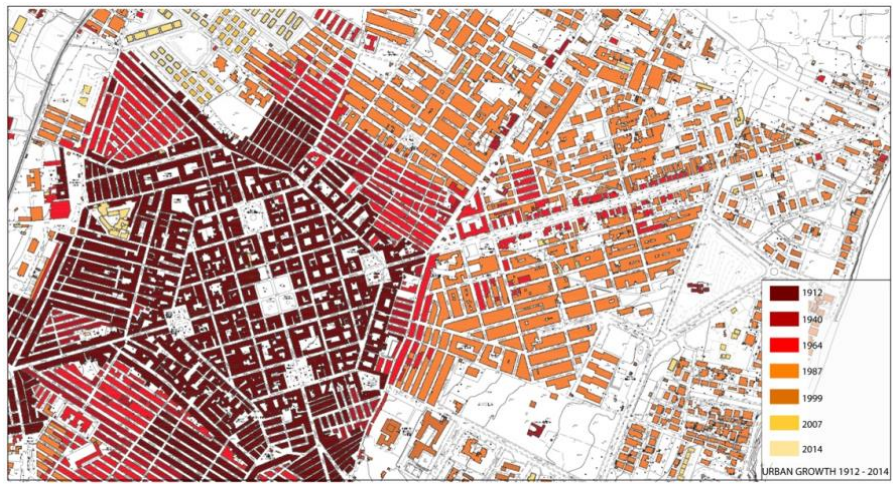

**Figure 3.** City of Avola—Urban Growth map.

Recently, the urban growth processes have been governed by poor quality urban plans that gives marginal attention to agricultural land protection and sustainability. The result are the new medium density settlements, developed close to the town centre, following an awkward interpretation of the modernist planning models [41].

During the new city masterplan design process, carried out from 2013 to 2016, several information, both geological and urbanistic, were collected and properly stored within geo-referenced digital data sets. The study, based on all the historical cartographies available, produced a map representing the growth of Avola settlement, from the foundation in the early 18th century to 2015. From this study the total number of buildings in this area, turned out to be $N_B$=17477. Historical cartographies of urban fabric were overlaid with new official cartography, released by Urban Planning Department of the Regional Government, in order to obtain a historical dating of the entire built up area. As a result, urban growth had been quantified and mapped measuring the built-up changes corresponding to seven dates (1912, 1940, 1964, 1987, 1999, 2007, 2014).

The resulting urban growth map gives, for each building of the urban fabric, the date in which it is present in the corresponding map. This allows an estimate of the period of construction for each building. In addition, using the data (height and surface) derived from the official vectorial cartography, the volume of each building of the urban fabric has been computed by using standard GIS functions. As a result, every building in Avola has been characterized by its volume, height and construction date in the GIS data set. In addition, the data set includes the same information for other buildings scattered in the territory around Avola. In particular, a square area with a side length of about 10 Km has been considered, as shown in Figure 4.

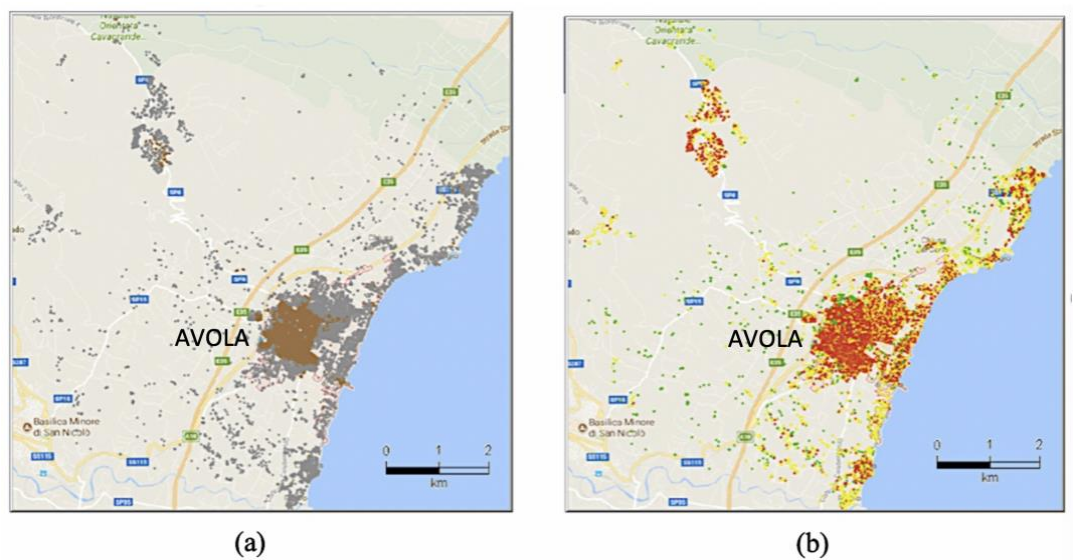

**Figure 4.** Territory of Avola. (**a**) Geographic Information System (GIS) data set: Masonry (brown) and reinforced concrete (grey) buildings; (**b**) initial vulnerability map: Low (green), medium (yellow) and high (red) vulnerability buildings.

Since the available data did not provide sufficient details on the construction typologies, we made a simplified assumption which is coherent with the aim of our methodological approach. In particular, all the buildings were classified in two main categories depending on their period of construction, as reported in Figure 4(a) with different colours: Masonry buildings (before 1965, in brown) and reinforced concrete buildings (after 1965, in grey). Then, crossing the construction information with data about the ratio between height H and base side L, and taking into account the minimum and maximum values of vulnerability allowed for the two considered structural typologies reported in Table 1, an initial vulnerability index $V_0$—see Equation (2)—has been assigned to each building following the prescriptions in Table 2.

**Table 2.** Rules for the assignment of the initial vulnerability index $V_0$.

| Building typology | Construction Date | Ratio H/L | $V_0$ is randomly assigned in the interval |
|---|---|---|---|
| Masonry | Before 1965 | < 0.5 | 0.3 - 0.7 |
| | | ≥ 0.5 and < 2.0 | 0.6 - 0.9 |
| | | ≥ 2.0 | 0.8 - 1.02 |
| Reinforced Concrete | ≥ 1965 and < 1988 | < 0.5 | 0.3 - 0.6 |
| | | ≥ 0.5 and < 2.0 | 0.55 - 0.85 |
| | | ≥ 2.0 | 0.8 - 1.02 |
| | ≥ 1988 and < 2008 | < 0.5 | 0.14 - 0.54 |
| | | ≥ 0.5 and < 2.0 | 0.5 - 0.8 |
| | | ≥ 2.0 | 0.7 - 1.02 |
| | ≥ 2008 | < 0.5 | -0.02 - 0.23 |
| | | ≥ 0.5 and < 2.0 | 0.2 - 0.55 |
| | | ≥ 2.0 | 0.5 - 0.7 |

In Figure 4(b) we represent in green buildings with low initial vulnerability ($-0.02 < V_0 < 0.3$), in yellow those with medium vulnerability ($0.3 < V_0 < 0.65$) and in red those with high vulnerability ($0.65 < V_0 < 1.02$).

### 2.4. Geological Framework and Corresponding GIS Data Set

Taking into account the above mentioned geo-referenced digital data sets, another "geology" layer has been used in the present analysis to empirically predict stratigraphic site-amplification (expressed by the previously introduced $c$ index), a factor which is primarily controlled by the thickness of soft sediments above a rigid substratum—see Figure 5(a). This stratigraphic configuration (soft/rigid) is quite common in the sector were the urban settlement of Avola extends since it consists of an ancient fluvial to marine depositional system discharging soft-sediments (clays, sands and conglomerates) which accumulated above a pre-existing (today buried) topographic surface modelled on carbonate (rigid) rocks.

The carbonate top-surface was reconstructed within the GIS environment by interpolating (Spline) the elevation of the top of the "rigid" geological formations, a point-value obtained from the consultation of a numbers of wells (and associated stratigraphic logs) available for the area. Thickness for the soft-sediments was therefore derived by subtracting the modern topographic surface (2 x 2 m cell size DTM, Digital Terrain Model) from the interpolated carbonate top-surface. Numerical values were then spatially joined to the polygon features describing the areal distribution of the outcropping geological formations. Since thickness can amplify or dampen the amplitude of seismic waves, a degree of amplification (medium-low, medium, medium-high and high) was associated (according to the thickness) to each geological formation, producing a new thematic map—see Figure 5(b), where the different geological areas are delimited by yellow lines tracked over a satellite image of the considered territory. With reference to Equation (4), expressing the earthquake intensity perceived by a given area, the following values for the amplification index $c$ can be assigned: $c = -0.5$ (for geological sites with a medium-low amplification), $c = 0$ (medium amplification), $c = 0.5$ (medium-high amplification), $c = 1$ (high amplification).

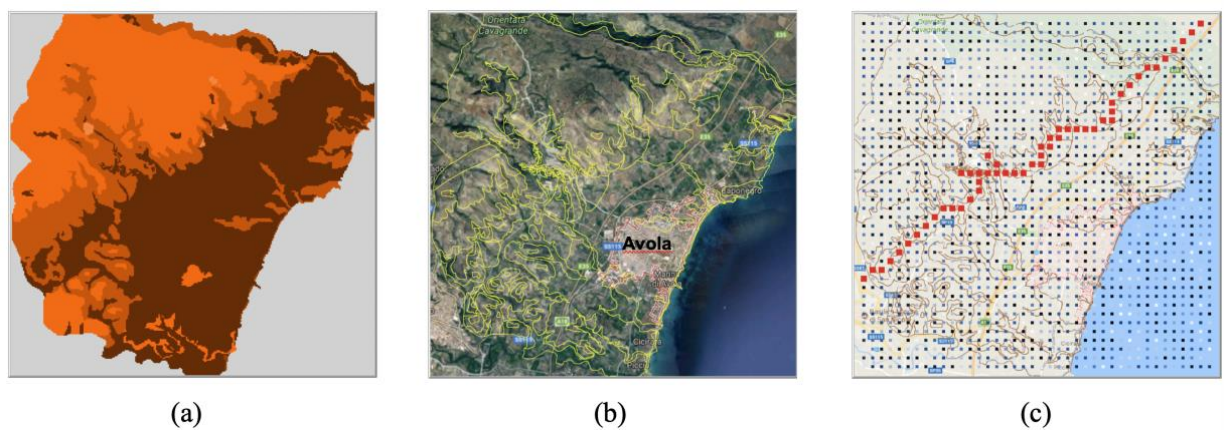

(a)                     (b)                     (c)

**Figure 5.** (**a**) Geological GIS data set; (**b**) site-amplification map of the territory of Avola; (**c**) OFC lattice where: Red nodes along the fault have a greater probability to be activated and to trigger an earthquake.

The territory of Avola is also sliced by a NE-SW trending tectonic structure which has been classified as an active and capable fault by ISPRA (Istituto Superiore per la Protezione e la Ricerca Ambientale) and added in the ITHACA (ITaly HAzard from CApable faults) data set (http://sgi2.isprambiente.it/ithacaweb/viewer). The tectonic structure consists of a 15 km-long, SE-dipping extensional fault located less than 3 km from the Avola city-center and separating the Avola mountains from the coastal plain with a 300 m-high morphological scarp. While its geometry at depth is still unknown, its occurrence gave us useful information in order to construct a realistic lattice taking into account that the triggering of an earthquake is more likely approaching the fault plane.

In Figure 5(c) the OFC small world lattice with N = 1600 nodes, already shown in Figure 2(a), is superimposed on the considered square map of the territory of Avola (links are hidden in the figure, only nodes are visible). In particular, the distance between two rows or two columns of the lattice corresponds to about 250 meters on the map. In order to incorporate the tectonic information in the OFC model, each node carries the value of the amplification index $c$ of the area immediately around it. Moreover, nodes along the fault (in red) also carry a seismic stress $F_i$ which—at variance with the other nodes—will take values in the interval [0.2, $F_{th}$]; therefore, these nodes will have a greater probability to be activated by the OFC dynamics and to trigger an earthquake.

### 2.5. Calibration of the Model

As already pointed out, the target of this study is to propose, by means of the new software ABES, an innovative methodology for evaluating the impact of a sequence of earthquakes on the vulnerability of the buildings present in the territory of Avola, under the assumption that this area, like the L'Aquila territory in 2009, would be in a critical state. To this purpose, the OFC earthquakes engine has to be finally integrated with the urban GIS data set within the ABES software, which needs also to be calibrated through a comparison with real data.

Considering an arbitrary sequence of seismic events, both the magnitude $M = ln\ S$ and the consequent intensity $I(M,c)$ are calculated in correspondence of each earthquake of size S. Then, all the $S$ active nodes of the OFC lattice transfer a seismic stress of intensity $I(M,c)$ to the buildings included into circles of radius R = (250 $m$ / 2)√2 centred on each of these nodes, as shown in Figure 6. Of course, buildings included in the overlapping areas (coloured in the figure) will be taken into account only once. In order to provide a realistic damage scenario in the considered urban area, it results that only a certain fraction $f_B$ of randomly chosen buildings inside each circle must be considered in the analyses, in order to take also into account the influence of possible external factors not present in the model. The value of this control parameter $f_B$ needs therefore to be calibrated by means of a comparison with real recorded data.

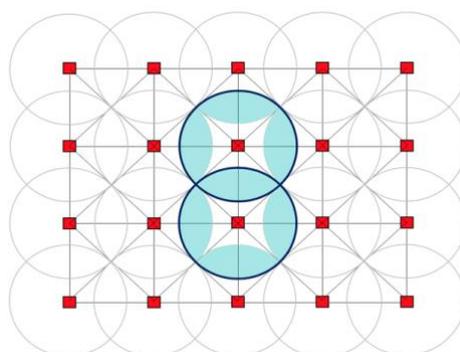

**Figure 6.** Overlapping circles around the active nodes. Inside these circles, only a fraction $f_B$ (control parameter) of buildings will be considered for the damage evaluation.

The idea is to reproduce, within our software, a seismic scenario approximately similar to the one occurred at L'Aquila in April 2009 and to compare the simulated effects of the virtual earthquakes on Avola's buildings with the real effects documented by several technical reports and investigations concerning the most destructive seismic sequence in Abruzzo [42–45]. In more detail, we consider the period of seismic activity going from 01/04/2009 to 10/04/2009: During these 10 days, hundreds of earthquakes occurred within the L'Aquila's territory (100 per day, in average), most of them of magnitude between 3 and 4 ML, with a mainshock of 5.9 ML and other three subsequent seismic events above 5 ML. The effects of this impressive seismic sequence on the urban structure of L'Aquila, and on the immediately surrounding areas, were evaluated in 17.40% of heavily damaged buildings (including buildings usable after intervention, partially unusable and temporarily unusable) and in 24.30% of destroyed (unusable) ones.

In the bottom panel of Figure 7 we report the size of a simulated seismic sequence of 1000 earthquakes (after a transient of 600 events) that, in some way, captures the main features of L'Aquila's earthquakes during the considered time interval of 10 days (with 100 events per day). The duration of 600 events of the transient state is common to all the simulations reported in the present paper and has been chosen simply by looking at the typical sequence shown in the top panel of Figure 2(b), since only after this transient the system enters into the critical state, where—as already observed—the average size of the earthquakes starts to increase and extreme events have a non-zero probability of occurrence. Notice also that each simulation run produces a different random sequence where the number and the intensity of extreme events cannot be controlled a priori, therefore the user can choose only a posteriori the sequence which better approximates the desired behaviour.

Looking to the magnitude of the 1000 events shown in Figure 7, we find 62 of them with 3ML < $M$ < 4ML, 11 events with 4ML < $M$ < 5ML and just three events with $M$ > 5ML. Among the latter events, all occurred during the last two days, we find a main shock with $M$ = 5.67ML followed by two other big peaks with, respectively, $M$ = 5.30ML and $M$ = 5.60ML. We can therefore assume this sequence as a good approximation of what really happened in the 10 days between 01/04/2009 and 10/04/2009 in L'Aquila territory and analyse the corresponding damage scenario for different values of the control parameter $f_B$.

In Table 3 we report the percentages of, respectively, heavily damaged and destroyed buildings for both the real seismic scenario of L'Aquila and the simulated one of Avola, the latter calculated by realizing different simulation runs for increasing values of $f_B$ (going from 0.1 to 1.0). As expected, the level of damage increases with $f_B$ and it results that the value $f_B$ = 0.4 produces the damage scenario most similar to the real one, for both the percentage of heavily damaged buildings (16.31%) and the percentage of destroyed buildings (24.33%).

**Table 3.** Comparison of the percentages of heavily damaged and destroyed buildings for the real seismic scenario of L'Aquila (Aprile 2009) and the simulated one of Avola (obtained for increasing values of the control parameter $f_B$).

| L'AQUILA 2009 (real data) | Heavily Damaged Buildings | Destroyed Buildings |
|---|---|---|
| | **17.40%** | **24.30%** |
| | | |
| AVOLA (simulations) | Heavily Damaged Buildings | Destroyed Buildings |
| $f_B$ = 0.1 | 7.75% | 4.22% |
| $f_B$ = 0.2 | 12.75% | 10.30% |
| $f_B$ = 0.3 | 15.16% | 17.34% |
| **$f_B$ = 0.4** | **16.31%** | **24.33%** |
| $f_B$ = 0.5 | 17.54% | 28.17% |
| $f_B$ = 0.6 | 16.89% | 34.44% |
| $f_B$ = 0.7 | 15.35% | 41.28% |
| $f_B$ = 0.8 | 13.57% | 47.90% |
| $f_B$ = 0.9 | 11.78% | 53.25% |
| $f_B$ = 1.0 | 10.51% | 55.00% |

In order to appreciate the progression of damages, the numbers of slightly damaged, heavily damaged and destroyed buildings (over a total of 17,477 buildings) are reported in the top panel of Figure 7 as function of the seismic sequence shown in the bottom panel. It is evident that the largest damage increment occurs in correspondence of the first main shock with $M$ = 5.67 ML, even if a further increase in the number of destroyed buildings can be also observed in correspondence of the last one of the three main events, with $M$ = 5.60 ML. It is worth to notice that the increase in the number of slightly damaged buildings is not necessarily monotonic, since sometimes—due to a single seismic event or to the cumulated effects of several events—a certain part of these buildings can change their status in heavily damaged, and the same holds for the number of heavily damaged buildings, since they can change their status in destroyed.

Of course, since several information about each single damaged or destroyed building are available (size, data of construction, vulnerability, typology, geological features of the edification soil, etc.), once calibrated the model this methodology also allows to perform any kind of statistical analysis of the progressive effects of different sequences of earthquakes in the critical state. In the next section we will address this point in the context of new hypothetical seismic scenarios involving the territory of Avola. Anyway, it is important to stress that, due to the lack of more detailed data on the building typologies, at the moment we will limit ourselves to show only the potentiality of our integrate approach, without pretending to give practical directions for urban planners and territory managers.

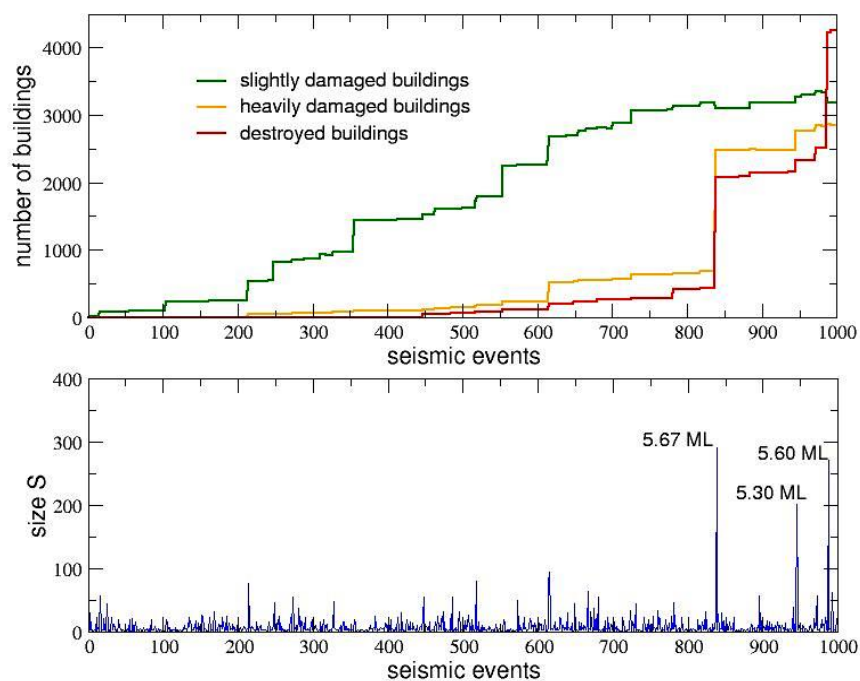

**Figure 7.** Bottom panel: A simulated sequence of 1000 seismic events (100 per day). Top panel: Evolution of the corresponding number of slightly damaged, heavily damaged and destroyed buildings for $f_B$ = 0.4.

## 3. Results and Discussion

Once calibrated the ABES software and fixed the control parameter $f_B$, in this section we simulate new seismic sequences of duration 10 days inside the critical state (again with 100 events per day, just after the usual transient of 600 events). All the seismic scenarios presented here are obtained running the simulations with different random realizations of the initial conditions (seismic stress distribution on the lattice nodes), in order to investigate how the increase of damage due to repetitive ground motions is influenced by both the buildings parameters and the geological features of the foundation soil. In particular, among the possible ones, we have chosen (a posteriori) three different seismic sequences: A "low intensity" seismic scenario, with no events of magnitude greater than 5ML; a "medium intensity" seismic scenario, having only one event of magnitude greater than 5ML; and a "high intensity" seismic scenario, characterized by the presence of two events of magnitude greater than 5ML.

### 3.1. Low Intensity Seismic Scenario

Figure 8 refers to the low intensity seismic scenario and plots, in its upper part, the increase in the total damage produced in the urban area by the 1000 seismic events of the chosen sequence, whose sizes are reported in the bottom part of the figure. In this scenario, only a few events have a magnitude included between 4 ML and 5 ML, and none of them exceeds 5 ML.

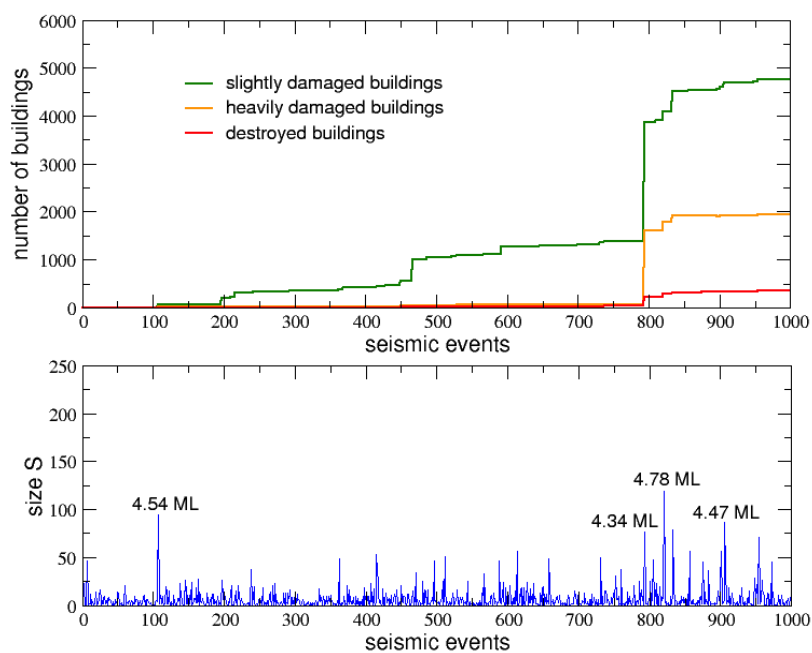

**Figure 8.** Low intensity seismic scenario: The number of (slightly or heavily) damaged and destroyed buildings are reported (top panel) as function of 1000 seismic events (bottom panel).

The total number of events (EV) registered up to days 1, 2, 8 and 10 are reported in Figure 9 for three different ranges of magnitude, together with the corresponding percentages of slightly damaged (SDB), highly damaged (HDB) and destroyed (DEB) buildings cumulated in the same time intervals.

In the last row of the table, the geographical damage distributions are plotted day by day (undamaged buildings are coloured in light green, slightly damaged in dark green, heavily damaged in yellow and destroyed in red). As it can be observed, at the end of the 10 days the majority of buildings remains undamaged by this low intensity seismic sequence, about 38% of them results to be either slightly or heavily damaged and only about 2% are destroyed. The main damage increase in this scenario occurs at day 8, when the overall number of damaged buildings jumps from less than 2% to more than 30% in correspondence of a single earthquake of magnitude 4.34 ML. The subsequent earthquakes of magnitude 4.78 ML and 4.47 do not make the situation much worse.

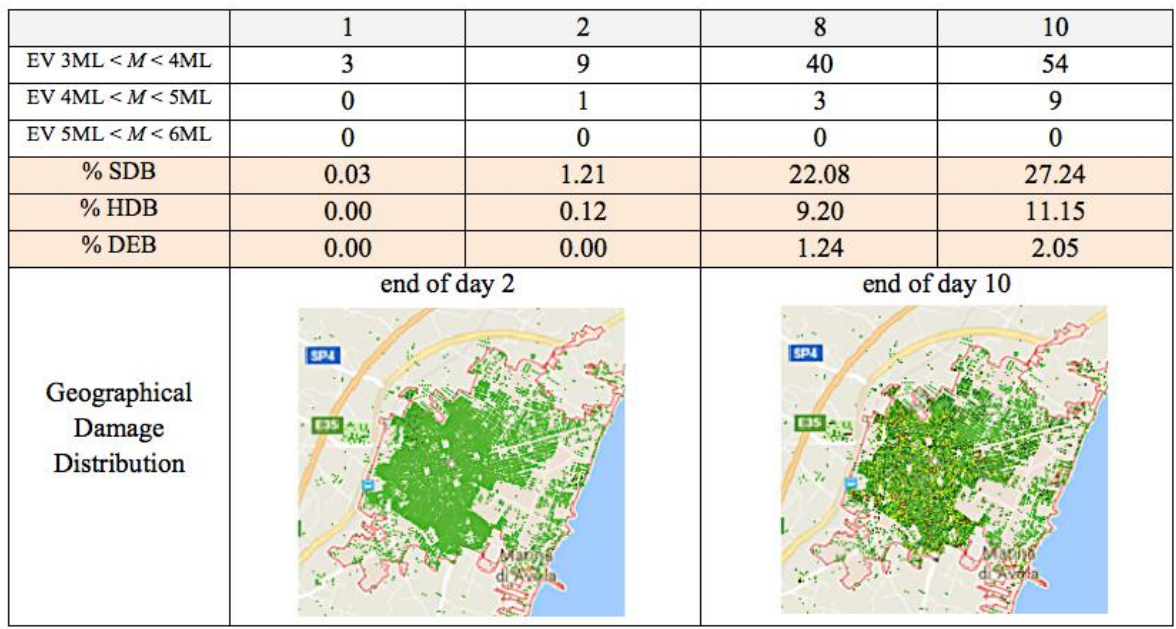

| | 1 | 2 | 8 | 10 |
|---|---|---|---|---|
| EV 3ML < $M$ < 4ML | 3 | 9 | 40 | 54 |
| EV 4ML < $M$ < 5ML | 0 | 1 | 3 | 9 |
| EV 5ML < $M$ < 6ML | 0 | 0 | 0 | 0 |
| % SDB | 0.03 | 1.21 | 22.08 | 27.24 |
| % HDB | 0.00 | 0.12 | 9.20 | 11.15 |
| % DEB | 0.00 | 0.00 | 1.24 | 2.05 |
| Geographical Damage Distribution | end of day 2 | | end of day 10 | |

**Figure 9.** Low intensity seismic scenario: Distribution of seismic events and percentages of slightly damaged (SDB), highly damaged (HDB) and destroyed (DEB) buildings within the considered time period.

### 3.2. Medium Intensity Seismic Scenario

The same study is repeated for the medium intensity seismic scenario, with a few events between 4 ML and 5 ML and only one event above 5 ML. The results are reported in Figures 10 and 11.

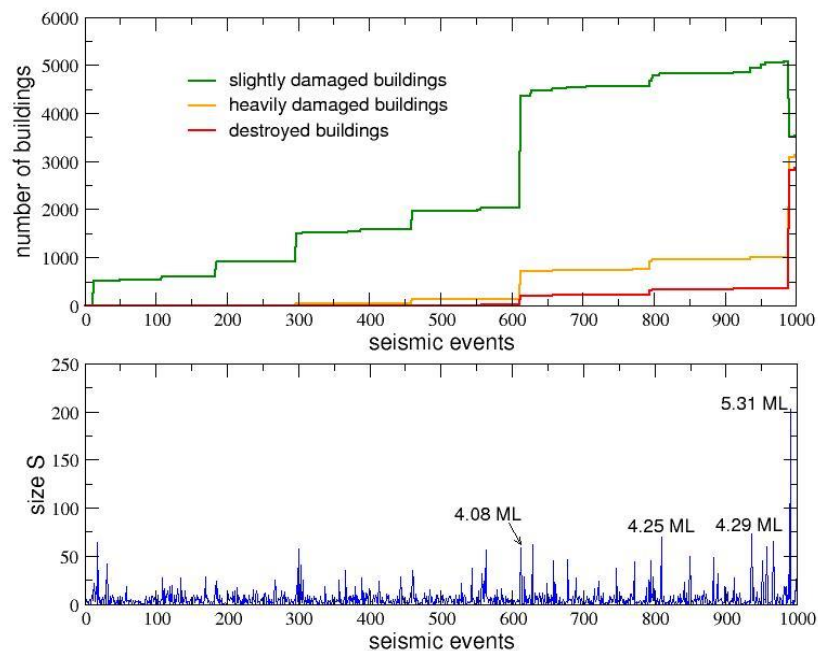

**Figure 10.** Medium intensity seismic scenario: The number of (slightly or heavily) damaged and destroyed buildings are reported (top panel) as function of 1000 seismic events (bottom panel).

Looking at the two panels of Figure 10, it clearly appears that the main increase in the percentage of slightly damaged buildings (from about 10% to more than 25%) occurs at the beginning of day 7, in correspondence of an earthquake of magnitude 4.08 ML, while the main increase in the percentage of destroyed buildings (from about 2% to about 16%) has been registered at the end of the last day, just after the main shock of magnitude 5.31ML. As expected, the percentages of highly damaged or destroyed buildings reported in Table 5 are sensibly greater than the correspondent ones in the previously studied low intensity scenario, as also confirmed by the spread of red spots on the geographical damage distribution shown in the last row of the table. It is worth to point out that the sudden decrease in the percentage of slightly damaged buildings, in correspondence of the last mainshock, is due to the fact that many of these buildings change their status in highly damaged or destroyed.

| DAY | 3 | 7 | 9 | 10 |
|---|---|---|---|---|
| EV 3ML < *M* < 4ML | 7 | 26 | 36 | 41 |
| EV 4ML < *M* < 5ML | 1 | 5 | 6 | 9 |
| EV 5ML < *M* < 6ML | 0 | 0 | 0 | 1 |
| % SDB | 8.53 | 25.96 | 27.67 | 20.21 |
| % HDB | 0.21 | 4.26 | 5.54 | 17.87 |
| % DEB | 0.00 | 1.29 | 1.97 | 16.37 |
| Geographical Damage Distribution | end of day 7 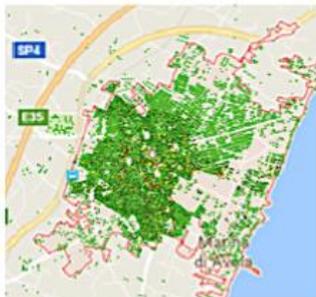 | | end of day 10 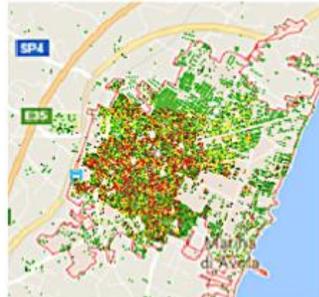 | |

**Figure 11.** Medium intensity seismic scenario: Distribution of seismic events and percentages of slightly damaged (SDB), highly damaged (HDB) and destroyed (DEB) buildings within the considered time period.

*3.3. High Intensity Seismic Scenario*

The last considered sequence of earthquakes simulates a high intensity seismic scenario, with many events between 4 ML and 5 ML, one event of 5.21 ML at day 7 and a second intense ground motion of 5.06 at day 9. The sequence is reported at the bottom of Figure 12 while in the upper part of the same figure the evolution of damage is shown, as usual, in terms of percentages of slightly damaged, heavily damaged and destroyed buildings.

Differently than what happened in the previously considered seismic sequences, in this case significant amounts of highly damaged and destroyed buildings can be observed. The values reported in Figure 13 reveal in fact, at the end of the considered time interval, the presence of about 46% of buildings either highly damaged or destroyed. Notice that the main jump in the percentage of slightly damaged buildings occurs in correspondence of an event of 4.33 ML, while the main increase in the percentage of heavily damaged buildings is observed in correspondence of the main shock of 5.21 ML. Finally, the percentage of destroyed buildings shows three main jumps, in correspondence of three events with magnitude 5.21 ML, 4.38 ML and 4.85 ML, respectively. It is interesting to highlight that—as it can be observed in some way also in the previous scenarios—the main damage jumps are not necessarily correlated to the main events: for example, the earthquake of 4.38 ML at the end of day 8 produces an increase of 10% of destroyed buildings while the stronger event of 5.06 ML at day 9 has much smaller effects (with an increase of only 2%).

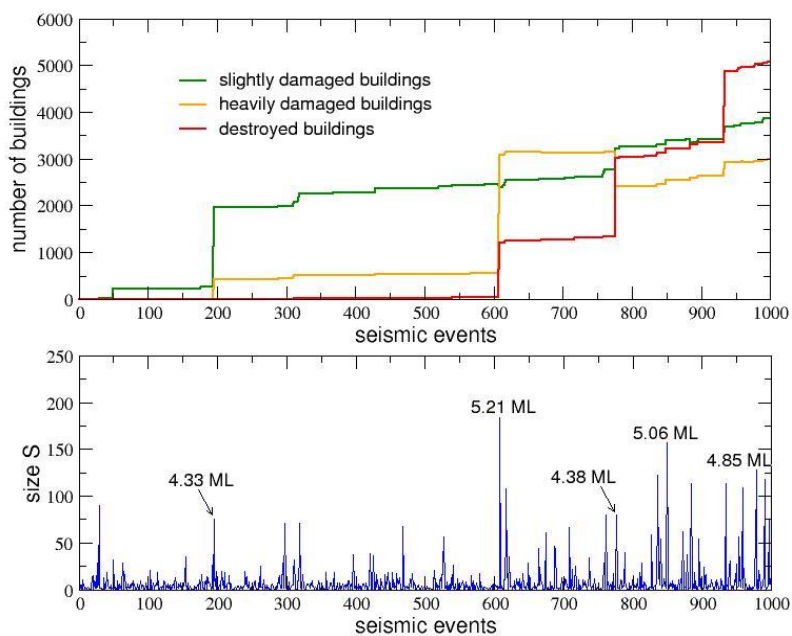

**Figure 12.** High intensity seismic scenario: The number of (slightly or heavily) damaged and destroyed buildings are reported (top panel) as function of 1000 seismic events (bottom panel).

This counterintuitive effect is due to the intrinsic non-linear nature of the damage accumulation in the real world, which is captured by the agent-based simulations thanks to the conjunction of three important features implemented in our software: The exponential increase of the energy released by an earthquake as function of its magnitude, the sigmoidal shape of the relation between the seismic intensity and the damage increment of a given building, and the step-like function describing the change of status of the building (which depends on two subsequent thresholds in its total cumulated damage). The combination of these features makes the overall system—like all the real complex systems—very sensitive to its past seismic history, in the sense that similar seismic events could have very different damage effects, not proportional to their intensity, just because they happen in different moments.

| | 2 | 7 | 8 | 10 |
|---|---|---|---|---|
| EV 3ML < *M* < 4ML | 6 | 23 | 28 | 40 |
| EV 4ML < *M* < 5ML | 2 | 9 | 12 | 23 |
| EV 5ML < *M* < 6ML | 0 | 1 | 1 | 2 |
| % SDB | 11.24 | 14.66 | 18.68 | 22.15 |
| % HDB | 2.47 | 17.94 | 13.86 | 17.08 |
| % DEB | 0.01 | 7.25 | 17.42 | 28.96 |
| Geographical Damage Distribution | end of day 7 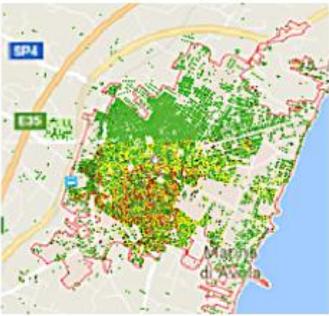 | | end of day 10 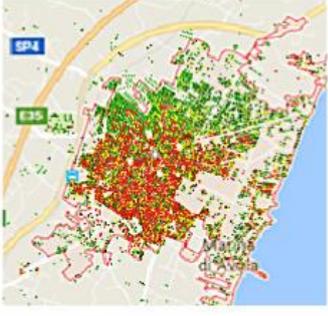 | |

**Figure 13.** High intensity seismic scenario: Distribution of seismic events and percentages of slightly damaged (SDB), highly damaged (HDB) and destroyed (DEB) buildings within the considered time period.

*3.4. Comparison of the Damages Produced in the Considered Seismic Scenarios*

In this section a more detailed statistical analysis of buildings' damages at the end of each simulation for the three previously considered seismic scenarios is presented.

Some characteristic parameters of each building have been taken into account, in particular its date of construction, the ratio H/L between the height of the building and the side length of its equivalent square plant, the initial vulnerability and the amplification level of the foundation soil.

This analysis could obviously be extended to many others characteristic parameters if more details on both the urban and geological GIS data sets would have been available.

The percentage of buildings belonging to each range of the considered parameters have been reported in the first column of Table 4. In the same table, in correspondence of every range of parameters, the percentages of buildings which, after each one of the three considered seismic sequences, result to be, respectively, undamaged (UND), slightly damaged (SDB), highly damaged (HDB) or destroyed (DEB), are reported.

**Table 4.** Influence of characteristic building's parameters on the damage intensity for the three considered seismic scenarios.

| Buildings' Parameters | | % of Buildings | Low Intensity Seismic Scenario | | | | Medium Intensity Seismic Scenario | | | | High Intensity Seismic Scenario | | | |
|---|---|---|---|---|---|---|---|---|---|---|---|---|---|---|
| | | | UND | SDB | HDB | DEB | UND | SDB | HDB | DEB | UND | SDB | HDB | DEB |
| Date of constr uction | 1912 - 1940 | 35.59 | 16.02 | 12.17 | 6.49 | 0.91 | 8.24 | 7.54 | 8.67 | 11.13 | 5.50 | 7.04 | 14.65 | 14.65 |
| | 1941 - 1964 | 14.42 | 7.68 | 4.13 | 2.15 | 0.46 | 4.74 | 3.22 | 3.63 | 2.83 | 2.59 | 2.83 | 5.38 | 5.38 |
| | 1965 - 1987 | 41.15 | 27.90 | 10.16 | 2.41 | 0.68 | 24.72 | 8.63 | 5.46 | 2.34 | 12.31 | 6.79 | 8.72 | 8.72 |
| | 1988 - 1999 | 2.90 | 2.26 | 0.54 | 0.09 | 0.01 | 2.24 | 0.55 | 0.06 | 0.05 | 0.83 | 0.29 | 0.17 | 0.17 |
| | 2000 - 2007 | 0.50 | 0.39 | 0.11 | 0.00 | 0.00 | 0.34 | 0.11 | 0.03 | 0.02 | 0.14 | 0.03 | 0.02 | 0.02 |
| | 2008 - 2018 | 5.44 | 5.29 | 0.14 | 0.00 | 0.00 | 5.17 | 0.26 | 0.01 | 0.00 | 0.77 | 0.10 | 0.02 | 0.02 |
| Ratio H/L | <0.5 | 24.49 | 21.13 | 3.28 | 0.07 | 0.01 | 18.94 | 4.42 | 1.01 | 0.12 | 7.87 | 2.04 | 1.29 | 1.29 |
| | 0.5-2 | 69.03 | 36.05 | 22.74 | 8.98 | 1.26 | 24.99 | 14.78 | 15.75 | 13.52 | 13.90 | 13.33 | 24.54 | 24.54 |
| | >2 | 6.48 | 2.36 | 1.22 | 2.10 | 0.80 | 1.62 | 1.01 | 1.11 | 2.73 | 0.37 | 1.72 | 3.13 | 3.13 |
| Initial vulner ability | low | 4.85 | 4.83 | 0.01 | 0.00 | 0.00 | 4.76 | 0.09 | 0.00 | 0.00 | 0.36 | 0.00 | 0.00 | 0.00 |
| | medium | 36.90 | 27.29 | 9.27 | 0.29 | 0.05 | 22.66 | 8.03 | 5.73 | 0.48 | 12.85 | 4.84 | 4.98 | 4.98 |
| | high | 58.25 | 27.41 | 17.97 | 10.86 | 2.01 | 17.92 | 12.25 | 12.19 | 15.89 | 9.03 | 12.28 | 23.79 | 23.79 |
| Site amplifi cation | medium-low | 6.85 | 6.12 | 0.71 | 0.03 | 0.00 | 5.12 | 1.46 | 0.27 | 0.01 | 1.90 | 0.54 | 0.51 | 0.51 |
| | medium | 8.26 | 6.83 | 1.40 | 0.03 | 0.00 | 6.11 | 1.36 | 0.77 | 0.02 | 3.22 | 0.75 | 0.37 | 0.37 |
| | medium-high | 0.07 | 46.51 | 25.14 | 11.09 | 2.06 | 0.07 | 0.00 | 0.00 | 0.00 | 0.01 | 0.00 | 0.01 | 0.01 |
| | high | 84.80 | 46.51 | 25.14 | 11.09 | 2.06 | 33.25 | 17.89 | 17.33 | 16.34 | 17.02 | 15.79 | 28.08 | 28.08 |

As it can be noticed, for the low intensity scenario the percentages of damaged buildings are small for all the considered parameters. Only two values are greater than 20% and refer in particular to slightly damaged buildings belonging to the range H/L between 0.5 and 2 or built on soils having high site amplification.

With reference to either the medium or high intensity seismic scenarios, it can be observed that the large majority of heavily damaged and destroyed buildings have been built before 1988. The influence of the ratio H/L on the presence of heavy damage or collapse is particularly significant in the range 0.5–2. The low values referred to ratios greater than 2 are related to the small presence of tall buildings in the considered area. With reference to the role of the initial vulnerability on the successive damage or collapse of the buildings, the results confirm that vulnerable structures are more prone to suffer severe damages. Finally, again as one could expect, the greatest percentages of heavily damaged and destroyed building are located on soils with high values of the site amplification parameter.

The comparison of the data related to the three seismic scenarios allows to point out that, for all the considered ranges of parameters, the percentages of undamaged buildings decreases with the intensity of the seismic scenario while the number of destroyed buildings increases.

The previous results show a satisfactory agreement with the expected damage scenarios for an urban area subjected to repetitive ground motions, thus further confirming the reliability of the proposed methodology.

## 4. Conclusions

This study represents a first attempt to apply a new multidisciplinary agent-based approach to the seismic assessment of a large urban, and peri-urban, area. By integrating competences and information coming from several scientific disciplines, going from the SOC dynamics of earthquakes to the seismic response of buildings with a given vulnerability, from the GIS features of the urban settlement to agent-based simulations, the proposed methodology allows to evaluate the effects of long sequences of seismic shocks on the buildings present in the area under investigation, assuming that the latter is in a critical state. In particular, differently than other simulation studies which refer to the effects produced on an urban area by a single seismic event, the present approach is able to reproduce the increase of buildings' damage due to repetitive ground motions. The proposed methodology has been implemented in a new software, called ABES, developed by the authors within the fully programmable multi-agent environment Netlogo.

In the paper, a small portion of the territory of Avola (Siracusa, Italy) has been considered as a case study. The similarity, from the point of view of the seismic risk, of this area with the one involved in the earthquake occurred in L'Aquila in 2009, allowed us to perform a calibration of the software based on real data. Then, among the many seismic scenarios that have been simulated for the considered territory (by randomly varying the initial conditions, i.e.; the seismic stress distribution on the lattice nodes), three different ones with an increasing number of strong events have been taken into account and discussed in the results section. The effects of the considered seismic sequences on the buildings of Avola have been analyzed in detail and the influence of some characteristic parameters on the damage has been evaluated.

In spite of the introduced simplified assumptions, due to the lack of sufficient details on the construction typologies and to the approximations at the basis of the SOC model, the numerical results clearly show the potentialities of the present approach in estimating the damage diffusion in the urban area due to repetitive ground motion.

It is important to highlight that the main goal of this study is not to provide a detailed report of the seismic vulnerability of the town of Avola but to propose a new methodology that could be applied to any geographic area for which both urban and geological Gis data sets are available. Obviously the more detailed the available data will be, the more reliable will turn out the results.

We believe that the proposed methodology could encourage forward-looking municipal administrations to adopt this kind of approach in order to implement both prevention and emergency plans concerning the related urban territory. In fact this approach could be used to investigate how to increase the resilience of urban areas, or to reduce the seismic vulnerability, by improving the structural performance of the most vulnerable buildings, through ad hoc retrofitting strategies. Furthermore, it could be used for planning new safety areas that could diminish the risk for the population, and for implementing more efficient and timely evacuation plans in case of repeated seismic events of either small or high intensity.

It is worth noting that the proposed methodology has been applied to the scale of a small town for investigating the distribution of the building damage intensity but it could also be easily extended to a larger scale in order to address the seismic vulnerability of several homogeneous urban areas interested by common seismic-genetic sources. For example, one could consider the case of the Oriental Sicily area, whose seismic risk is mainly associated to the Ibleo-Maltese system of faults that were responsible for the great devastating 1963 earthquake.

~~**Supplementary Materials:** The following are available online at www.mdpi.com/xxx/s1, Figure S1: title, Table S1: title, Video S1: title.~~

**Author Contributions:** Conceptualization, Annalisa Greco, Alessandro Pluchino, Andrea Rapisarda and Ivo Caliò; Methodology, Annalisa Greco and Alessandro Pluchino; Software, Alessandro Pluchino; Validation, Annalisa Greco and Alessandro Pluchino; Investigation, Annalisa Greco and Alessandro Pluchino; Resources, Luca Barbarossa, Giovanni Barreca, Francesco Martinico; Data Curation, Luca Barbarossa, Giovanni Barreca, Francesco Martinico; Writing-Original Draft Preparation, Annalisa Greco and Alessandro Pluchino;; Writing-Review & Editing, Annalisa Greco, Alessandro Pluchino and Andrea Rapisarda; Visualization, Annalisa Greco and Alessandro Pluchino; Supervision, Annalisa Greco, Alessandro Pluchino, Andrea Rapisarda and Ivo Caliò; Funding Acquisition, Annalisa Greco, Alessandro Pluchino and Andrea Rapisarda

**Funding:** This research was partly funded by the PRIN 2017WZFTZP "Stochastic forecasting in complex systems" and partly by the University of Catania, with the project "Linea di intervento 2" of the Department of Physics and Astronomy "Ettore Majorana" and the project "Piano della Ricerca Dipartimentale 2016-2018" of the Department of Civil Engineering and Architecture.

**Acknowledgments:** Authors would like to thank Vittorio Rosato and Sonia Giovinazzi for their useful suggestions and comments.

**Conflicts of Interest:** The authors declare no conflict of interest. The funders had no role in the design of the study; in the collection, analyses, or interpretation of data; in the writing of the manuscript, or in the decision to publish the results.